\documentclass[%
 reprint,
 amsmath,amssymb,
 aps,
pra,
floatfix,
longbibliography,
]{revtex4-1}

\usepackage{graphicx}% Include figure files
\usepackage{dcolumn}% Align table columns on decimal point
\usepackage{bm}% bold math
\usepackage{multirow}
\usepackage{array}

\begin{document}

\preprint{APS/123-QED}

\title{Superscattering from Subwavelength Corrugated Cylinders}
%\thanks{A footnote to the article title}%

\author{Vitalii I. Shcherbinin$^{1,2}$}
\author{Volodymyr I. Fesenko$^{1,3}$}  
\author{Tetiana~I.~Tkachova$^{2}$}
\author{Vladimir~R.~Tuz$^{1,3}$}
 \email{tvr@jlu.edu.cn; tvr@rian.kharkov.ua}
\affiliation{$^1$State Key Laboratory of Integrated Optoelectronics, College of Electronic Science and Engineering, International Center of Future Science, Jilin University, 2699 Qianjin Street, Changchun 130012, China}
\affiliation{$^2$National Science Center `Kharkiv Institute of Physics and Technology', National Academy of Sciences of Ukraine, 1, Akademicheskaya Street, Kharkiv 61108, Ukraine}
\affiliation{$^3$Institute of Radio Astronomy, National Academy of Sciences of Ukraine, 4 Mystetstv Street, Kharkiv 61002, Ukraine}

\date{\today}

\begin{abstract}
Wave scattering from a cylinder with a tensor impedance surface is investigated based on the Lorentz-Mie theory. A practical example of such a cylinder is a subwavelength metallic rod with helical dielectric-filled corrugations. The investigation is performed with the aim to maximize scattering cross-section by tailoring the surface impedance of cylindrical scatterers. For the normally incident TE$_z$ and TM$_z$ waves the required surface impedance of a subwavelength cylinder can be produced by longitudinal (axial) and transverse (circumferential) corrugations, respectively. It is shown that such corrugations induce superscattering at multiple frequencies, which can be widely tuned with either or both the size and permittivity of dielectric-filled corrugations. In the microwave band, this effect is demonstrated to be robust to material losses and is validated against the full-wave simulations and experiment. For the TE$_z$ waves the enhanced scattering from the cylinder is found to have a broad frequency bandwidth, provided that the relative permittivity of corrugations is low or equal unity. In the latter case, the corrugated cylinder acts as an all-metal superscatterer. For such cylinders the near-field measurements are implemented and provide the first experimental evidence of the superscattering phenomenon for all-metal objects. In addition to multifrequency superscattering, the dielectric-filled corrugations are shown to provide multifrequency cloaking of the cylinder under the incidence of the TM$_z$ waves. Simultaneous superscattering and cloaking at multiple frequencies distinguishes corrugated cylinders from other known practicable scatterers for potential applications in antenna designing, sensing, and energy harvesting.
\end{abstract}

%\pacs{41.20.Jb, 42.25.Bs, 78.67.Pt}
% 41.20.Jb	Electromagnetic wave propagation; radiowave propagation 
% 42.25.Bs	Wave propagation, transmission and absorption           
% 78.67.Pt	Multilayers; superlattices; photonic structures; metamaterials 

\maketitle
%\tableofcontents
Enhancement of wave scattering from small objects is a vital issue in present-day technologies \cite{vercruysse_2013, ng_2013, zeng_2014, smith_2015, zhang_2015, han_2015, gholipour_2017}, including miniaturized antennas, sensors and energy harvesting devices. This issue is directly related to the natural constraint inherent in most subwavelength scatterers \cite{alu_2005}. This constraint is known as a single-channel limit \cite{ruan_2010}, which represents an upper limit to scattering cross-section for such scatterers and is attained under resonance condition for one of the scattering modes (channels). The only way to overcome this constraint is to ensure resonant scattering of several modes at a single frequency. Such a resonance overlapping magnifies scattering from a given object and is known as superscattering \cite{ruan_2010, ruan_2011}. The larger the number of resonant modes, the larger the superscattering cross-section. Therefore, theoretically, superscattering opens a way to arbitrary enhancement of wave scattering from subwavelength objects. In practice, however, this effect is often hindered by the lack of low-loss materials and appropriate design solutions.

In the infrared and visible parts of spectrum, the superscattering can be realized in  cylindrical structures formed by several plasmonic and dielectric layers. In the pioneering work \cite{ruan_2010}, for such a structure the total scattering cross-section was optimized to exceed the single-channel limit by a factor of eight. However, in actual conditions, the enhancement of wave scattering from this scatterer appears to be four times lower \cite{ruan_2010} or even disappears \cite{mirzaei_2013}. The reason is losses in plasmonic material (metal). Similar effect was also reported in Ref.~\cite{mirzaei_2013} for a core-shell plasmonic nanowire, which exhibits scattering cross-section in slight excess of the single-channel limit in the presence of material losses. The situation is much the same for terahertz frequencies. In this frequency band, the plasmonic component of superscatterers can be replaced by conducting materials such as graphene \cite{li_2016, raad_2019} or hexagonal boron nitride (BN) \cite{qian_2018}, which can initiate resonance overlapping for different scattering modes. However, losses in these materials notably degrade the performance of such scatterers, making efficient superscattering practically elusive in the terahertz band. The effect of losses becomes of minor importance for good conductors (metals) in the microwave band. Their use as microwave superscatterers, however, is hampered by the fact that metals on their own do not support any scattering resonances due to small surface impedance. The required low-loss surface impedance can be realized in all-metal or metal-dielectric metasurfaces \cite{padooru_2012}. Such metasurfaces are usually made in the form of periodic arrays of subwavelength elements (strips, crosses, patches, etc. \cite{Radi_PhysRevApplied_2015}), which can be fabricated by patterning a metallic layer. The advantageous use of metasurfaces composed of circumferential metallic strips was recently demonstrated by the first experimental evidence of superscattering in Ref.~\cite{qian_2019}, in which measured scattering cross-section was reported to be about five times the single-channel limit. This impressive result offers great prospect for practical use of superscatterers in the microwave band.

The experimental findings of Ref.~\cite{qian_2019} are related to scattering of the TM$_z$ waves (the TE waves in the notation of Ref.~\cite{qian_2019}) from multi-layered metasurface-dielectric cylinder. Note that multi-layered structures made of alternating plasmonic-dielectric \cite{ruan_2010, ruan_2011, mirzaei_2013, mirzaei_2014}, graphene-dielectric \cite{li_2016, raad_2019}, BN-dielectric \cite{qian_2018}, dielectric-dielectric \cite{liu_2017} or metasurface-dielectric \cite{qian_2019} layers represent the most common design solution for cylindrical superscatterers. This raises the following questions: Is it possible to realize a superscatterer of simpler design with a single surface (metasurface)? Is it possible to achieve superscattering without dielectric? Is it possible to design an efficient superscatterer for the TE$_z$ waves in the microwave band? Our answer is yes. All this is possible with a subwavelength corrugated cylinder. It has long been known that such a cylinder is characterized by the averaged surface impedance \cite{Barlow_1954, piefke_1959, harvey_1960, davies_1962, katsenelenbaum_1966, clarricoats_1969}, which may induce scattering resonances associated with the so-called spoof (or designer) surface plasmons (see Ref.~\cite{huidobro_2018} and references therein). In this paper, it is shown that properly-designed corrugations ensure overlapping of these resonances, resulting in superscattering.

\section{Scattering from a tensor impedance cylinder}
\label{sec:tensor}

\begin{figure}[t!]
\centerline{\includegraphics[width=1.0\linewidth]{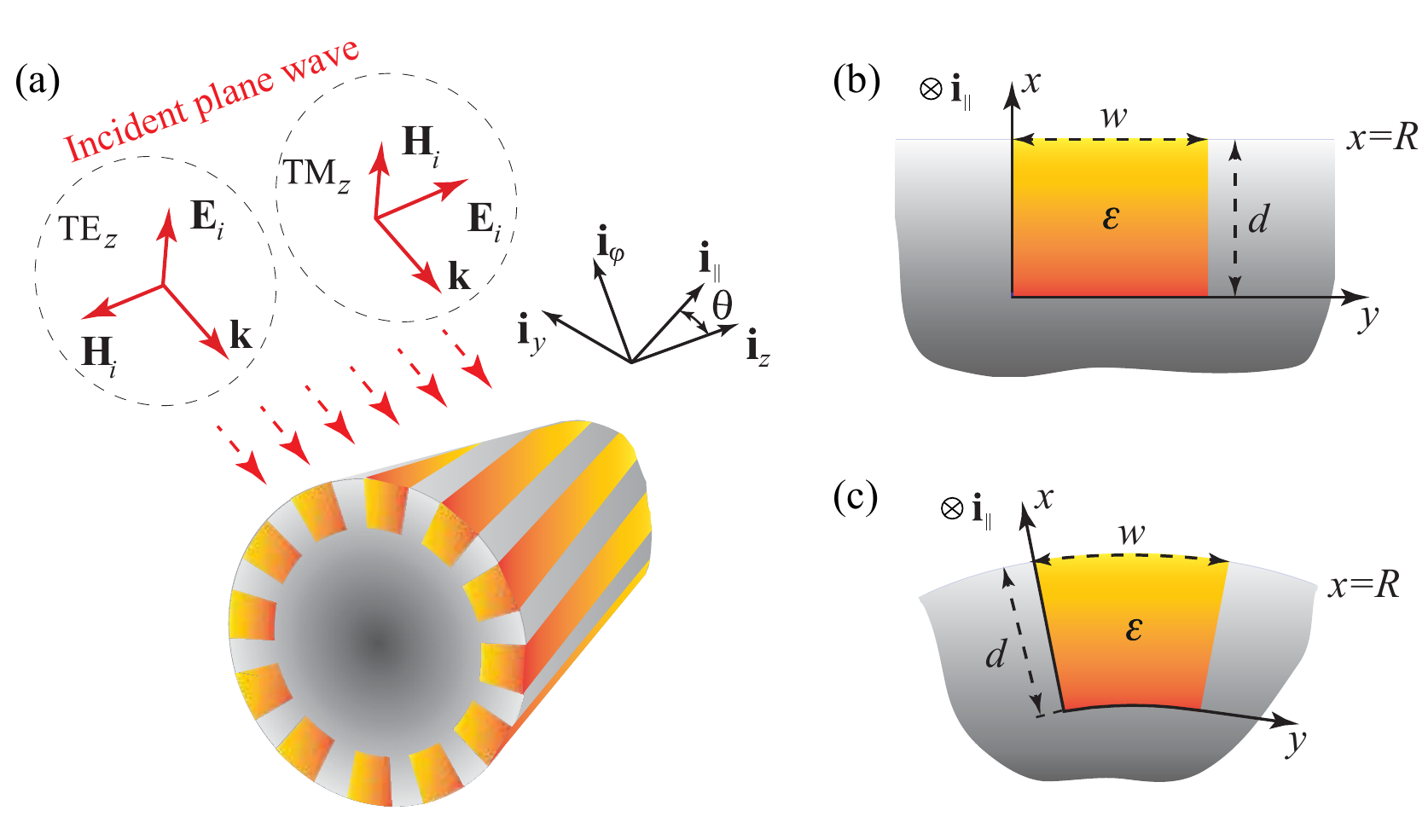}}
\caption{(a) Metallic cylinder with helical dielectric-filled corrugations, and structure of (b) rectangular and (c) wedge-shaped corrugations.} \label{fig:fig1}
\end{figure}

Consider an arbitrary-polarized plane wave propagated in a space with relative permittivity $\varepsilon_2$ and permeability $\mu_2=1$. The wave has the frequency $\omega$ and is normally incident on a cylinder of radius $R$. In cylindrical coordinates $\{r, \varphi, z\}$, the total field of incident and scattered waves can be expanded in terms of azimuthal modes as \cite{bohren_2008}
\begin{equation}
\begin{split}
&\left\{ \begin{matrix}
   {H_{z2}} \cr
   {E_{z2}} \cr
\end{matrix} \right\}\\
& = \sum^{\infty}_{n=-\infty} i^n H_0 F_n \left( \left\{ \begin{matrix}
   {1} \cr
   {P} \cr
\end{matrix}\right\}
J_n(k_2r) + \left\{ \begin{matrix}
   {a_n} \cr
   {b_n} \cr
\end{matrix}\right\} 
H_n^{(1)}(k_2r)\right),
\end{split}\label{eq:azimuthal}
\end{equation}
where $F_n=\exp(-i\omega t+i n\varphi)$, $J_n(\cdot)$ is the Bessel function, $H_n^{(1)}(\cdot)$ is the Hankel function of the first kind, $n$ is the azimuth index, which numerates scattering modes (channels), $k_2=\sqrt{\varepsilon_2}k$, $k=\omega/c$, $\{a_n,b_n\}$ are the dimensionless amplitudes of scattered wave, $H_{zi}=H_0=[0.5(1+Q)]^{1/2}$ and $E_{zi}=PH_0=[0.5(1-Q)]^{1/2}\exp(i\beta)$ are the amplitudes of incident wave, $Q\in [-1,1]$, $\beta\in[-\pi,\pi]$, $Q$ defines polarization of incident wave and equals 1, $-1$ and 0 for the TE$_z$-, TM$_z$- and dual-polarized waves [see Fig.~\ref{fig:fig1}(a)], respectively.

The unknowns $a_n$ and $b_n$ are determined by the boundary conditions on the cylinder surface. For generality, assume that the cylindrical scatterer features a tensor surface impedance such that 
\begin{equation}
\begin{matrix}
   {E_{\varphi 2}} = Z_{\varphi \varphi}H_{z 2} - Z_{\varphi z} H_{\varphi 2}, \cr
   {E_{z 2}} = -Z_{zz}H_{\varphi 2} + Z_{z \varphi} H_{z 2}, \cr
\end{matrix} \label{eq:impedance}
\end{equation}
for $r=R$.

In this case, substitution of Eq.~(\ref{eq:azimuthal}) into Eq.~(\ref{eq:impedance}) gives the following explicit expressions for the scattering coefficients:
\begin{equation}
\begin{split}
a_n = &-\left[\left(\lambda_1 + Z_{\varphi \varphi} - \varepsilon_2 P Z_{\varphi z}\lambda_1\right)\left(1+\varepsilon_2 Z_{zz}\lambda_2\right)\right. \\&- \left.\varepsilon_2 Z_{\varphi z}\lambda_2\left(P+\varepsilon_2 P Z_{zz}\lambda_1 - Z_{z\varphi}\right)\right] \lambda_3 D^{-1},\\
b_n = &-\left[\left(\lambda_2+Z_{\varphi \varphi}\right)\left(P + \varepsilon_2 P Z_{zz}\lambda_1 - Z_{z\varphi}\right)\right. \\ &+ \left.Z_{z \varphi}\left(\lambda_1 + Z_{\varphi \varphi} - \varepsilon_2 P Z_{\varphi z}\lambda_1\right)\right]\lambda_3 D^{-1}, \\
\end{split} \label{eq:coeff_1}
\end{equation}
where $D = \left(\lambda_2 + Z_{\varphi \varphi}\right)\left(1 + \varepsilon_2 Z_{zz}\lambda_2\right) - \varepsilon_2 Z_{\varphi z} Z_{z \varphi}\lambda_2$, $\lambda_1 = ikk_2^{-1}J^\prime_n(k_2R)/J_n(k_2R)$, $\lambda_2 = ikk_2^{-1}H^{\prime(1)}_n(k_2R)/H^{(1)}_n(k_2R)$, and $\lambda_3=J_n(k_2 R)/H_n^{(1)}(k_2 R)$.

With the amplitudes $a_n$ and $b_n$, one obtains the scattering efficiency \cite{bohren_2008}
\begin{equation}
Q_\textrm{sca} = \frac{2}{kR}N_\textrm{sca} = \frac{2}{kR\left(1+|P|^2\right)} \sum_{n=-\infty}^\infty\left(|a_n|^2 + |b_n|^2 \right),
\label{eq:qsca}
\end{equation}
where $N_\textrm{sca}$ is the scattering cross-section normalized to the single-channel scattering limit $2\lambda/\pi$ \cite{alu_2005, ruan_2010}. Both $Q_\textrm{sca}$ and $N_\textrm{sca}$ serve as a measure of wave scattering from an impedance cylinder of radius $R$.

For clarity sake, it is convenient to rewrite Eq.~(\ref{eq:coeff_1}) in the following form \cite{alu_2005, alu_2005_2}:
\begin{equation}
a_n=\frac{U_n^{\textrm{TE}}}{U_n^{\textrm{TE}}+iV_n^{\textrm{TE}}},~~~b_n=\frac{U_n^{\textrm{TM}}}{U_n^{\textrm{TM}}+iV_n^{\textrm{TM}}},
\label{eq:coeff_2}
\end{equation}
where $U_n$ and $V_n$ are the real-valued functions in the lossless (and gainless) case. 

In this case, the scattering coefficients are bounded quantities, which attain the peak values $|a_n|=1$ and $|b_n|=|P|$ at resonance $V_n=0$ \cite{alu_2005}. Under resonance condition, the scattering cross-section of the $n$-th azimuthal mode reaches the single-channel limit  \cite{ruan_2010}, which corresponds to the highest possible level of wave scattering for the most subwavelength objects. A way to overcome this limit for small scatterers is to ensure resonance overlapping for two or more scattering channels. This phenomenon is known as superscattering  \cite{ruan_2010, ruan_2011}. In a similar manner, it is possible to achieve cloaking of a subwavelength object, provided that the conditions $U_n=0$ are satisfied for dominant (low-order) scattering modes  \cite{padooru_2012, alu_2005_2, alu_2009, Fleury_PhysRevApplied_2015, fesenko_2018, shcherbinin_2019}. It should be noted that, despite limitation on the scattering cross-section, resonant objects of a smaller size can scatter waves more efficiently than nonresonant scatterers of the same dimensions [see Eq.~(\ref{eq:qsca})]. The widely-known example of a highly efficient scatterer is an atom at resonant frequency \cite{ruan_2010}.

Theoretically, a tensor impedance surface furnishes the most general way to manipulate scattering from the object under the incidence of arbitrary-polarized waves. However, in modern practice, there are only few ways to design and tailor such a surface (metasurface). A known way is to introduce a patterned metallic coating on a dielectric substrate \cite{fong_2010, patel_2013, quarfoth_2015}. In the cylindrical geometry, example is a dielectric wire coated by helical conducting strips \cite{shcherbinin_2018, shcherbinin_2019}. An alternative design solution is described in Appendix \ref{sec:cylinder}. It is in the form of a subwavelength metallic rod with helical periodic corrugations [Fig.~\ref{fig:fig1}(a)].

One of the means of manipulating wave scattering from a cylinder can be provided by off-diagonal components $Z_{\varphi z}$ and $Z_{z \varphi}$  of the surface impedance tensor. This is because these components give rise to a cross-polarization coupling [see Eq.~(\ref{eq:coeff_1})] and thereby enable polarization conversion of waves normally incident on the cylinder. In other words, tensor impedance metasurfaces make it possible to manipulate polarization of scattered waves \cite{fong_2010, selvanayagam_2014} and thereby may advance functionality of the scatterer. Although the problem of polarization control by metasurfaces is of much current interest \cite{wu_2019, mun_2019, akram_2019, wu_2019_2}, it deserves detailed consideration, which is beyond the scope of the present investigation. In the following, we are aimed at enhancing scattering of the TE$_z$ and TM$_z$ waves from a cylinder by means of surface impedance, which induces no cross-polarization coupling. To attain this goal, we take advantage of longitudinal wedge-shaped and transverse ring-shaped corrugations (see Appendix \ref{sec:cylinder} for more details). The corrugations in hand have the period $p$, the width $w$, the depth $d$, and are filled completely with dielectric of permittivity $\varepsilon$.

\section{Superscattering of the TE$_z$ waves}
\label{sec:super_te}

First, we consider the TE$_z$ wave $(Q=1)$ normally incident on a perfectly electric conducting (PEC) cylinder with longitudinal ($\bf{i}_\parallel\parallel\bf{i}_z$) wedge-shaped corrugations [Fig.~\ref{fig:fig1}(c)] described by an averaged anisotropic surface impedance with a single non-zero component
\begin{equation}
Z_{\varphi\varphi} = -\frac{w}{p}\frac{ik}{k_\bot}\frac{J^\prime_0(k_\bot R)-BN^\prime_0(k_\bot R)}{J_0(k_\bot R)-BN_0(k_\bot R)},
\label{eq:impedance_ff}
\end{equation}
where $k_\bot=\sqrt{\varepsilon}k$, $B=J^\prime_0(k_\bot R_d)/N^\prime_0(k_\bot R_d)$, $R_d=R-d$, $N_m(\cdot)$ denotes the $m$-th order Neumann function (for derivation of Eq.~(\ref{eq:impedance_ff}) see Appendix~\ref{sec:cylinder}). The aim is to investigate the ability of such corrugations to enhance wave scattering from the cylinder in free space ($\varepsilon_2=1$, $\mu_2=1$). 

\begin{figure*}
\centerline{\includegraphics[width=1.0\linewidth]{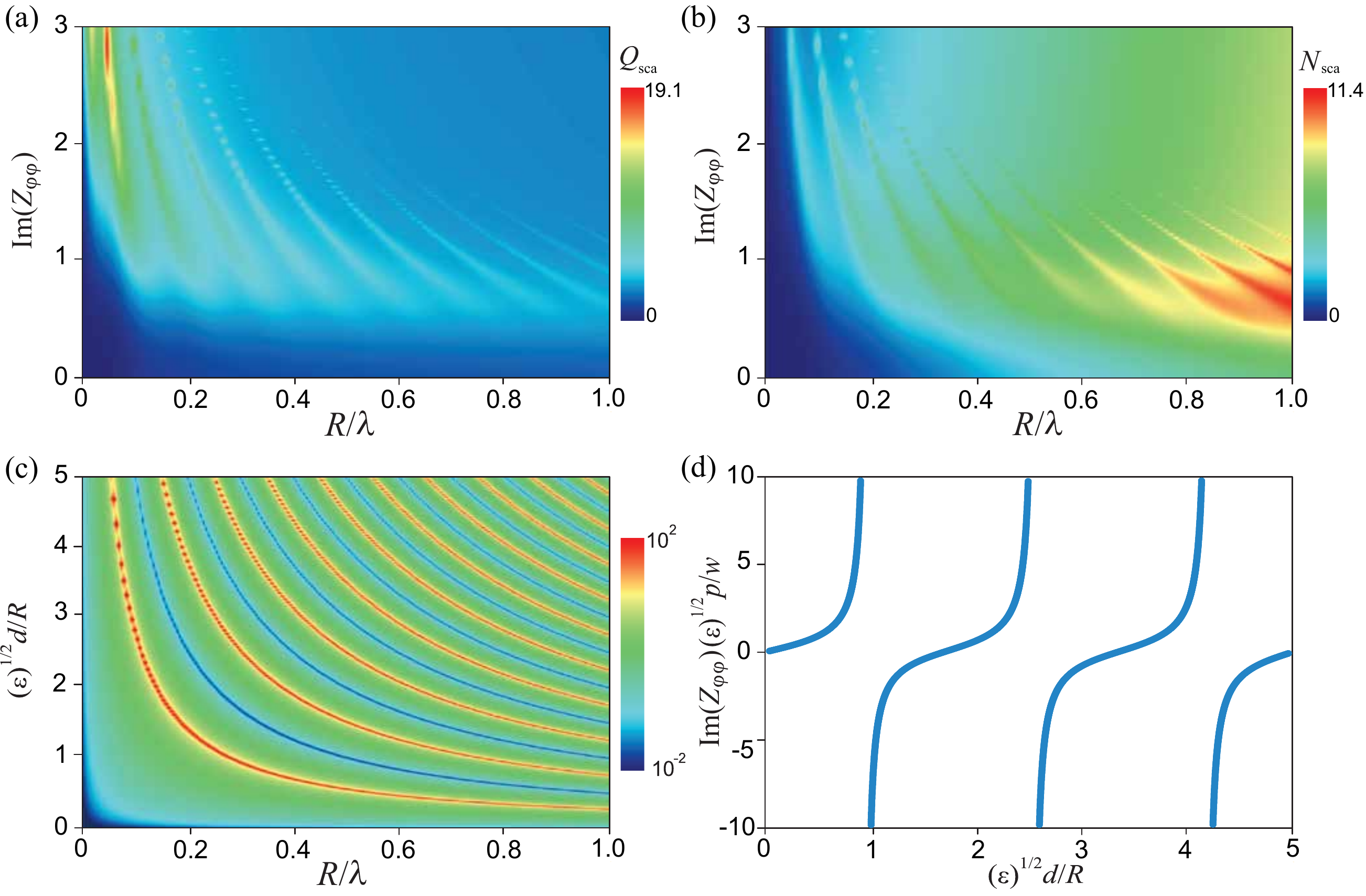}}
\caption{(a) Scattering efficiency and (b) normalized scattering cross-section of the TE$_z$ wave normally incident on a cylinder with radius $R$ and anisotropic surface impedance $Z_{\varphi\varphi}$ and $Z_{zz}=0$; (c) absolute value, and (d) imaginary part $(R/\lambda=0.3)$ of normalized surface impedance $Z_{\varphi\varphi}\sqrt{\varepsilon}p/w$ versus parameters of corrugations.} \label{fig:fig2}
\end{figure*}

Figures~\ref{fig:fig2}(a) and \ref{fig:fig2}(b) show the scattering efficiency $Q_\textrm{sca}$ and normalized scattering cross-section $N_\textrm{sca}$ as functions of radius $R$ and surface impedance $Z_{\varphi\varphi}$ of the subwavelength cylinder. It can be seen that the smaller the scatterer radius, the higher is the value of $\textrm{Im}Z_{\varphi\varphi}$ required to attain the maximum of $Q_\textrm{sca}$. Besides, smaller corrugated cylinders scatter waves more efficiently than smooth PEC rods ($Z_{\varphi\varphi}=0$) of the same radius and, in this regards, can be considered as so-called electromagnetic ``meta-atoms'' \cite{ruan_2010}. However, the total scattering cross-section of such ``meta-atoms'' falls bellow the single-channel limit ($N_\textrm{sca}=1$) for very small $R$ [Fig.~\ref{fig:fig2}(b)]. Thus the radius of cylinder with constant anisotropic surface impedance $Z_{\varphi\varphi}$ and $Z_{zz}=0$ must be selected large enough to achieve distinct contribution from two or more scattering channels. Figure~\ref{fig:fig2}(b) provides the design value of surface impedance required to maximize scattering cross-section of the subwavelength scatterer. In theory, this surface impedance can be of any origin. In practice, it can be realized by corrugating the surface of a metallic rod.

Figure~\ref{fig:fig2}(c) depicts the normalized absolute value of $Z_{\varphi\varphi}$ [see Eq.~(\ref{eq:impedance_ff})] as a function of corrugation parameters $\varepsilon$, $w/p$, $d$ and $R$. Since $d<R$, this figure indicates a lower limit on permittivity $\varepsilon$, which provides the desired impedance $Z_{\varphi\varphi}$ for cylindrical superscatterer of radius $R$. It is evident that this limit generally increases as $R$ decreases. In Fig.~\ref{fig:fig2}(c) one can distinguish a number of extrema of $Z_{\varphi\varphi}$. This suggests that a single corrugated metasurface may enhance scattering from desired cylinder at several frequencies as does a multilayered coating \cite{raad_2019, qian_2019}. As Fig.~\ref{fig:fig2}(a) illustrates, for such frequencies $Z_{\varphi\varphi}$  should be near the peak values shown in Fig.~\ref{fig:fig2}(c). In these peaks, the surface impedance $Z_{\varphi\varphi}$ in fact changes the sign [Fig.~\ref{fig:fig2}(d)]. Thus longitudinal wedge-shaped corrugations can provide both positive and negative values of $\textrm{Im}Z_{\varphi\varphi}$. From Fig.~\ref{fig:fig2}(d) one can notice that the higher is the design value of surface impedance, the closer should be the fabrication tolerances for corrugations. It is also clear that it is necessary to keep the permittivity $\varepsilon$ of corrugations as low as possible in order to alleviate this technological constraint. Our simulations additionally show that wave scattering from a corrugated cylinder is slightly sensitive to loss tangent $(\tan\delta)$ of dielectrics, provided that $\varepsilon$ is low. As indicated in Figs.~\ref{fig:fig2}(c) and \ref{fig:fig2}(d), the lower permittivity $\varepsilon$, the larger are ratios $d/R$ and $w/p$ required to attain the high averaged impedance $Z_{\varphi\varphi}$ of a corrugated surface. In our simulations, we set the width-to-period ratio $w/p$ of corrugations equal $0.9$, unless otherwise stated.

\begin{figure*}
\centerline{\includegraphics[width=1.0\linewidth]{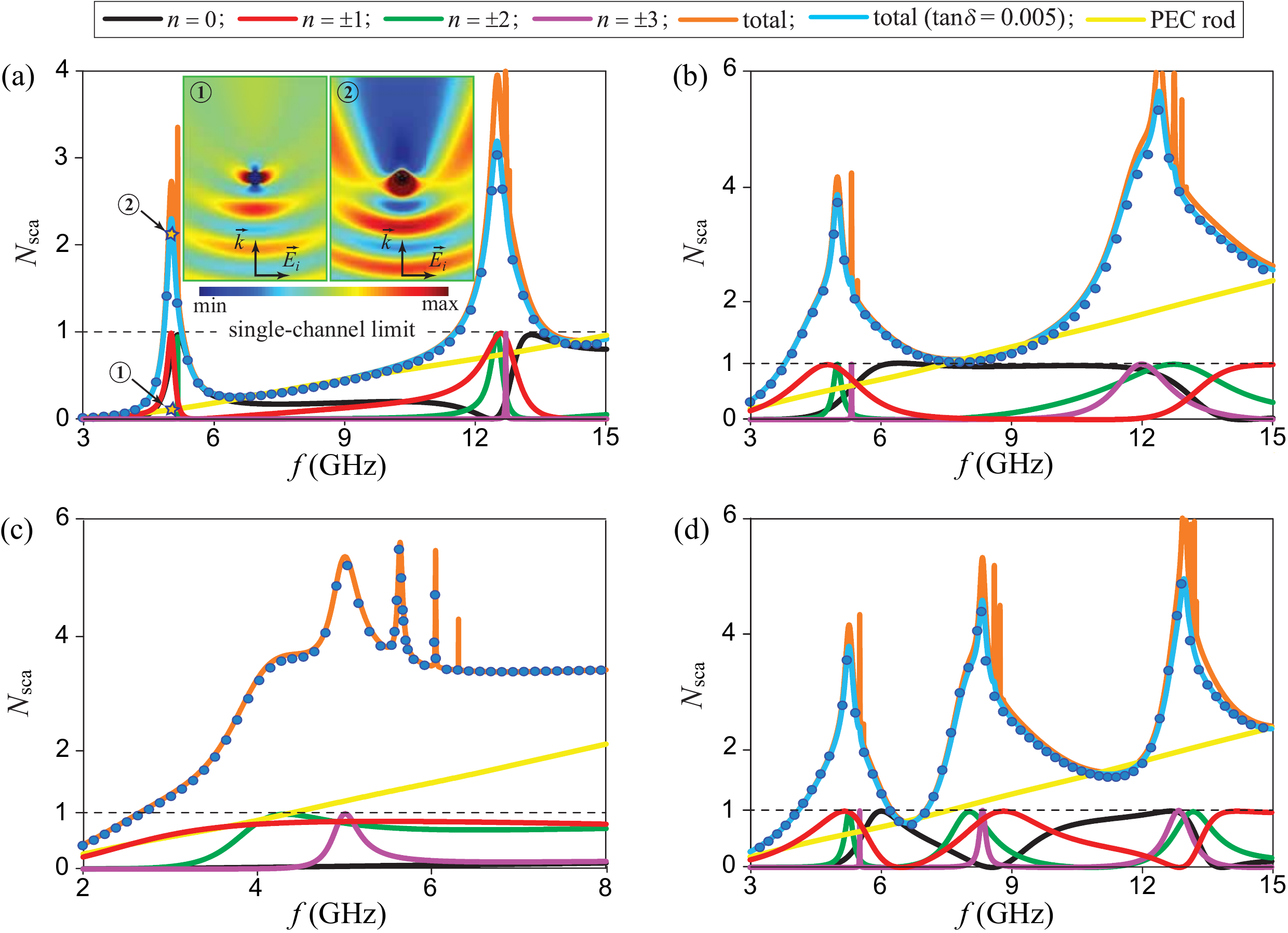}}
\caption{Total and partial scattering cross-sections of the TE$_z$ wave versus frequency for a corrugated cylinder with (a) $R=0.5$~cm, $d=0.39$~cm, and $\varepsilon=22$; (b) $R=1.0$~cm, $d=0.92$~cm, and $\varepsilon=4$; (c) $R=d=1.61$~cm and $\varepsilon=1$. Case (d) corresponds to wave scattering from a cylinder of radius $R=1$~cm with two distinctly-sized corrugations per period $p$, each characterized by the same values of $w/p=0.45$ and $\varepsilon=4$, and different depth of $0.92$~cm and $0.46$~cm. Scattering cross-section of PEC rod of the corresponding radius is shown by the yellow line and results of full-wave simulations are indicated by circular markers.} \label{fig:fig3}
\end{figure*}

In these simulations, for definiteness the frequency of the microwave scatterers of different radii is taken to be the same and equals 5~GHz. This frequency will be called the operating frequency.  Figure~\ref{fig:fig2}(c) indicates the required parameters of corrugations, which maximizes both $Q_\textrm{sca}$ [Fig.~\ref{fig:fig2}(a)] and $N_\textrm{sca}$  [Fig.~\ref{fig:fig2}(b)] for the subwavelength cylinder at the operating frequency. For $R=0.5$~cm $(R/\lambda\approx 0.08)$ the resultant scattering cross-section $N_\textrm{sca}$ versus frequency is shown in Fig.~\ref{fig:fig3}(a). In this case, permittivity $\varepsilon$ of corrugations is set equal to 22. Such permittivity is typical for low-loss $(\tan\delta\le 0.005)$ ceramics used in designing microwave metasurfaces \cite{xu_2019, sayanskiy_2019, fesenko_2019}. From Fig.~\ref{fig:fig3}(a) one can notice that for the operating frequency the scattering cross-sections from both $n=0$ and $n=\pm 1$ channels are close to the single-channel limit and therefore result in enhanced wave scattering (superscattering). An important point is that this effect is robust against dielectric losses, as opposed to superscattering induced by $n=0$ and $n=\pm2$ channels for nearby frequency of $5.2$~GHz. As a result, even with losses, for the operating frequency the scattering cross-section of the corrugated cylinder is more than twenty times as large as that of the PEC rod of the same radius $R=0.5$~cm. Moreover, as noted above, corrugations may provide multifrequency superscattering from a cylinder. This phenomenon is also clearly seen from Fig.~\ref{fig:fig3}(a), which demonstrates scattering enhancement simultaneously for frequencies of $5$ and $12.5$ GHz. For $12.5$ GHz the superscattering comes from interference of azimuth modes with $n=\pm 1$ and $n=\pm 2$.

A similar situation takes place for a cylinder of larger radius $R = 1$~cm $(R/\lambda\approx0.17)$. In this case, permittivity $\varepsilon$ of corrugations can be reduced up to 4 (e.g., as for an epoxy resin). As Fig.~\ref{fig:fig3}(b) illustrates, such permittivity ensures superscattering from the cylinder for the operating frequency of $5$~GHz and larger frequency of $12.4$~GHz as a result of resonance overlapping for modes with indices $n=\pm 1, \pm 2$ and $n=0, \pm 2, \pm 3$, respectively. From comparison of Figs.~\ref{fig:fig3}(a) and \ref{fig:fig3}(b) one can notice that, along with robustness to material losses, low-index dielectrics benefit from wideband superscattering. In particular, Fig.~\ref{fig:fig3}(b) demonstrates enhanced scattering from PEC rod with relative bandwidth in excess of 120\% with respect to the operating frequency. 

What is more important, superscattering from a corrugated cylinder can be achieved without invoking dielectrics. This is shown in Fig.~\ref{fig:fig3}(c) and, to our knowledge, demonstrates the first data reported for an all-metal superscatterer. Note that such all-metal superscatterer is relatively lightweight, since its geometrical transverse cross-section is smaller than that of a metal rod of the same radius.

It is also worth noting that a multifrequency superscattering from a corrugated cylinder is not a particular problem. It can be easily resolved by filling neighboring corrugations with different dielectrics. Alternatively, this can be done by introducing two or more distinctly-sized corrugations within one period $p$ of the structure. Figure~\ref{fig:fig3}(d) shows this, as an example, for a doubly-corrugated cylinder of radius $R=1$ cm. 

The phenomenon of superscattering from a corrugated cylinder has been validated against full-wave simulations by RF module of COMSOL Multiphysics\textsuperscript{\textregistered} solver. In simulations, the corrugation period $p$ is taken to satisfy the condition $p/\lambda<0.03$ and thus is small enough to avoid coupling of spatial harmonics due to corrugations \cite{tkachova_2019}. The results of full-wave simulations are shown in Fig.~\ref{fig:fig3} by markers and agree closely with our analytical findings. Patterns of the electric near-field ($|{\bf E}|^2$), which are plotted in the insets of Fig.~\ref{fig:fig3}(a) for corrugated and smooth rods at the operating frequency, serve as an additional evidence in favour of enhanced scattering of the TE$_z$ waves from a PEC cylinder with longitudinal wedge-shaped corrugations.

\section{Enhanced and reduced scattering of the TM$_z$ waves}
\label{sec:super_tm}

\begin{figure*}
\centerline{\includegraphics[width=1.0\linewidth]{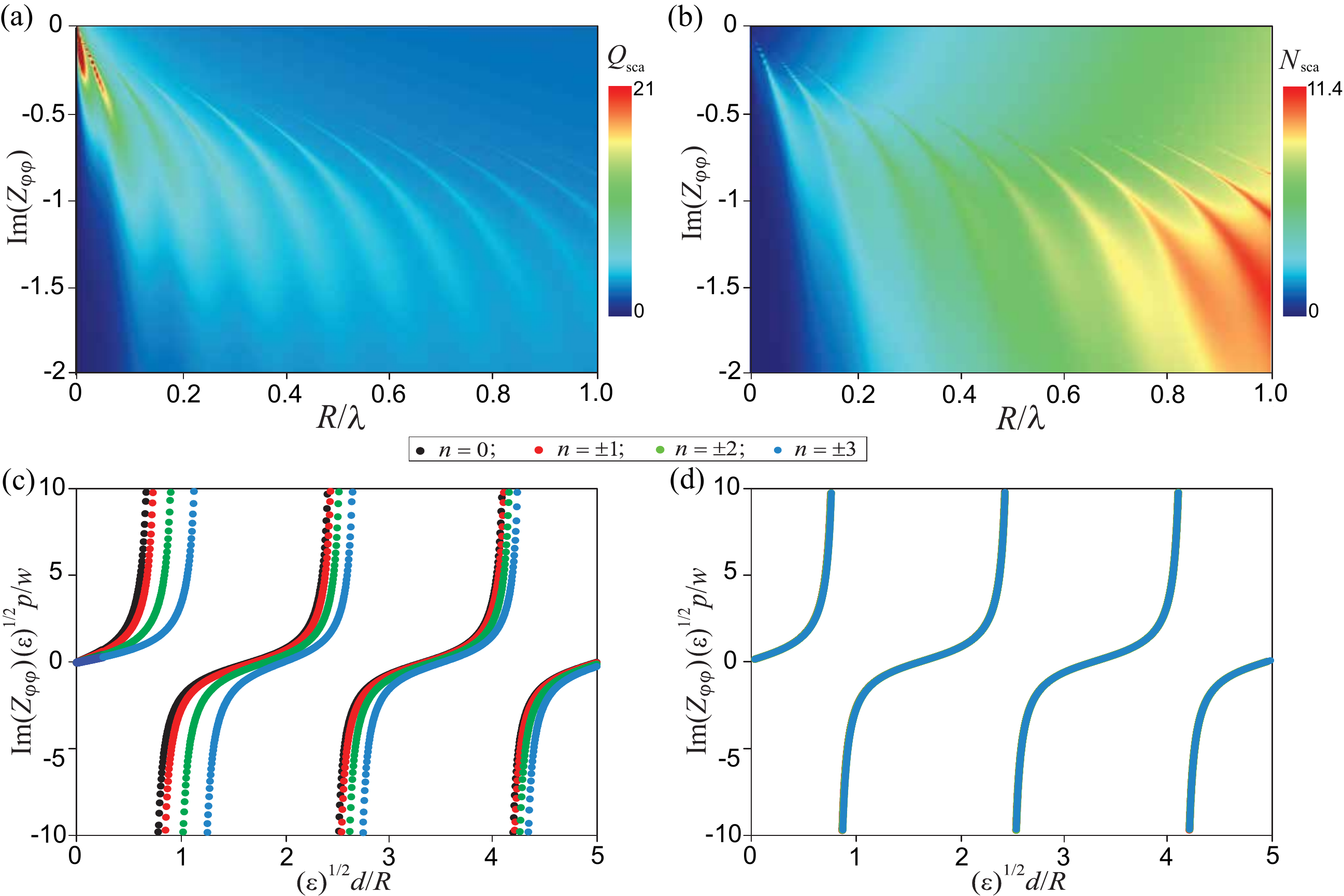}}
\caption{(a) Scattering efficiency and (b) normalized scattering cross-section of the TM$_z$-polarized wave normally incident on cylinder with radius $R$ and anisotropic surface impedance $Z_{\varphi\varphi}=0$ and $Z_{zz}$, and imaginary part of normalized surface impedance $Z_{zz}\sqrt{\varepsilon}p/w$ as a function of azimuth index $n$ and parameters of corrugated cylinder with $R/\lambda=0.3$ for (c) $d/R=0.5$ and (d) $d/R=0.1$.} \label{fig:fig4}
\end{figure*}

To enhance scattering of the TM$_z$ waves $(Q=-1)$ from a PEC cylinder we now turn our attention to transverse ($\bf{i}_\parallel\parallel\bf{i}_\varphi$) ring-shaped corrugations characterized by the averaged surface impedance with the nonzero component
\begin{equation}
Z_{zz} = \frac{w}{p}\frac{ik_\bot}{\varepsilon k}\frac{J_n(k_\bot R)-CN_n(k_\bot R)}{J^\prime_n(k_\bot R)-CN^\prime_n(k_\bot R)},
\label{eq:impedance_zz}
\end{equation}
where $C=J_n(k_\bot R_d)/N_n(k_\bot R_d)$ (see Eq.~(\ref{eq:imped_zz}) in Appendix~\ref{sec:cylinder}).

Figures~\ref{fig:fig4}(a) and \ref{fig:fig4}(b) show the scattering efficiency and normalized scattering cross-section of subwavelength cylinder versus radius $R$ and constant (mode-independent) surface impedance $Z_{zz}$. Again, one can notice that impedance cylinders of smaller radii scatter waves more efficiently than the PEC rods $(Z_{zz}=Z_{\varphi\varphi}=0)$, but at the same time should be large enough to overcome the singe-channel scattering limit. As is seen from Figs.~\ref{fig:fig4}(a) and \ref{fig:fig4}(b), the imaginary part of the surface impedance must be negative to enhance scattering of the TM$_z$ waves from the cylinder. Such a values of $Z_{zz}$ can be provided by corrugations. However, the results shown in Figs.~\ref{fig:fig4}(a) and \ref{fig:fig4}(b) should be generally considered as an estimation for a corrugated cylinder. This is because for such cylinder the surface impedance (\ref{eq:impedance_zz}) is not constant, but depends on the azimuth index $n$.

\begin{figure}[t!]
\centerline{\includegraphics[width=1.0\linewidth]{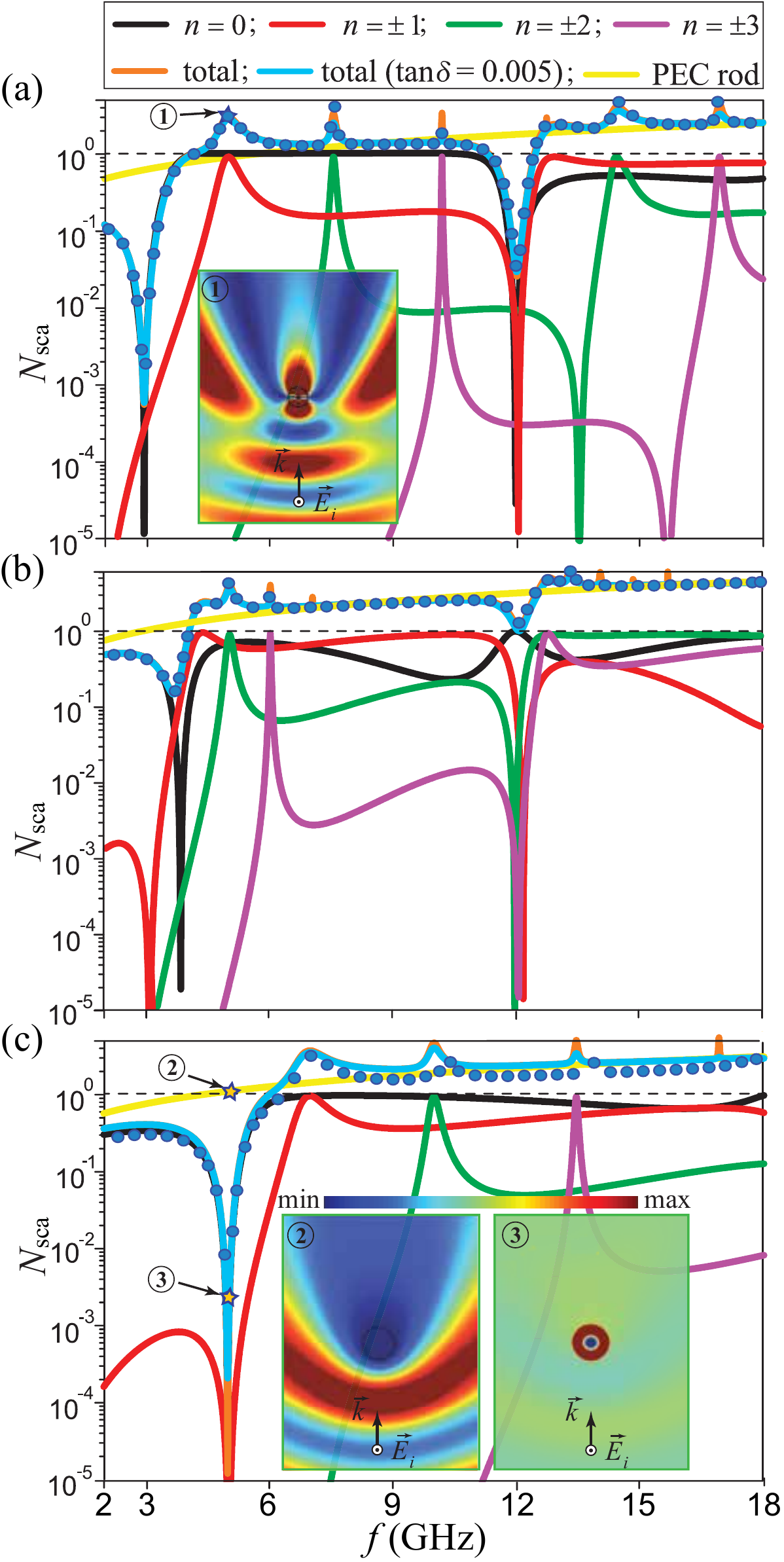}}
\caption{Total and partial scattering cross-sections of the TM$_z$ wave versus frequency for corrugated cylinder with (a) and (c) $R=0.5$~cm, $d=0.4$~cm, and $\varepsilon=22$; (b) $R=1.0$~cm, $d=0.4$~cm, and $\varepsilon=22$; (d)  $R=0.5$~cm, $d=0.36$~cm, $w/p=0.47$, and $\varepsilon=11$, along with results for PEC rods of the same radii (yellow lines) and data of full-wave simulations (circular markers).} \label{fig:fig5}
\end{figure}

For different $n$ the dependence of $Z_{zz}$ on the corrugation parameters $w/p$, $\varepsilon$, $d$, and $R$ has much in common with that shown in Fig.~\ref{fig:fig2}(c) for $Z_{\varphi\varphi}$. However, the larger the ratio $d/R$, the sharper is the distinction between the values of $Z_{zz}$ for different $n$. This distinction can be clearly seen from Fig.~\ref{fig:fig4}(c) in the case of $d/R=0.5$ and $R/\lambda=0.3$. By contrast, for small $d/R$ the impedance $Z_{zz}$ depends only slightly on $n$. An illustrative example is furnished by Fig.~\ref{fig:fig4}(d) for $d/R=0.1$ and $R/\lambda=0.3$. In this case, for low-order modes $Z_{zz}$ in Eq.~(\ref{eq:impedance_zz}) behave like a constant surface impedance.

In simulations, for cylinders of radii $0.5$~cm and $1.0$~cm we set $w/p=0.9$ and $\varepsilon=22$ and then optimize the corrugation depth in order to maximize scattering cross-section of the TM$_z$-polarized wave at the operating frequency of $5$~GHz. The resulting dependence of scattering cross-section $N_\textrm{sca}$ on frequency is shown in Figs.~\ref{fig:fig5}(a) and \ref{fig:fig5}(b). As in the case of the TE$_z$ waves, constructive interference of two or more scattering modes results in the multifrequency superscattering of the TM$_z$ waves from a corrugated cylinder. It is remarkable that for the operating frequency this effect is only slightly affected by dielectric losses, as indicated by Figs.~\ref{fig:fig5}(a) and \ref{fig:fig5}(b). 

Although our investigation is mainly concerned with the phenomenon of superscattering, we have also revealed that corrugations may additionally provide cloaking of a subwavelength PEC cylinder under the incidence of the TM$_z$ waves. Similar effect of reduced and enhanced scattering from a single structure was reported in Refs.~\cite{mirzaei_2013} and \cite{alu_2005_2} for core-shell plasmonic nanowire and sphere (see Fig.~6 in Ref.~\cite{alu_2005_2}), respectively. For the corrugated cylinder of radius 0.5 cm the reduced scattering can be seen from Fig.~\ref{fig:fig5}(a). It is notable that in this case cylinder cloaking by corrugations occurs simultaneously for two frequencies close to $3$ and $12$ GHz. Thus the PEC cylinder with transverse ring-shaped corrugations possesses a unique feature of both cloaking and superscattering at multiple frequencies. For the operating frequency of $5$ GHz the reduced scattering from a cylinder can be achieved by optimization of corrugation parameters. In the case of $R=0.5$~cm, the results of such optimization are shown in Fig.~\ref{fig:fig5}(c). It can be seen that in this case highly efficient cloaking of a cylindrical scatterer occurs owing to simultaneous suppression of the fundamental $(n=0)$ and the first $(n=\pm 1)$ scattering modes by corrugations. Thus surface corrugating represents a new cloaking technique for metallic cylinders.

As Fig.~\ref{fig:fig5} indicates, our simple analytical approach is in good agreement with full-wave simulations and therefore is particularly suitable for fast and accurate designing of corrugated cylindrical scatterers. From this figure it is apparent that subwavelength corrugations offer considerable scope for manipulation of wave scattering. This can also be seen from the near-field structure of the TM$_z$ wave in the insets of Figs.~\ref{fig:fig5}(a) and \ref{fig:fig5}(c), which show both superscattering and cloaking properties for corrugated cylinders of identical radius at the operating frequency.

\section{All-metal superscatterer}
\label{sec:allmet}

Recent experimental observations report superscattering for the TM$_z$ waves \cite{qian_2019}. Here, the purpose is to demonstrate the same phenomenon for the TE$_z$ waves. For this purpose, we make use of the simple design solution in the form of an all-metal superscatterer. Two prototypes of the superscatterer have been designed and fabricated. For the sample preparation cylindrical copper billets with a diameter of 3.0 cm and a length of 40 cm were used. A longitudinal wedge-shaped corrugations were engraved on the billets using a precise milling machine. The design parameters of the corrugations of the first (second) scatterers are as follows: $N=36$ ($N=41$), $w=0.15$ cm ($w=0.65$ cm) and $d=0.7$ cm ($d=0.5$ cm). According to our analytical estimations, these parameters ensure the operating frequencies of about 8.46 GHz $(R/\lambda\approx 0.42)$ and 10.68 GHz $(R/\lambda\approx 0.52)$ for the all-metal superscatterers with $N=36$ and $N=41$, respectively. These frequencies are within the frequency range $3-15$ GHz, which is covered by our experimental facilities. 

\begin{figure*}[t!]
\centerline{\includegraphics[width=1.0\linewidth]{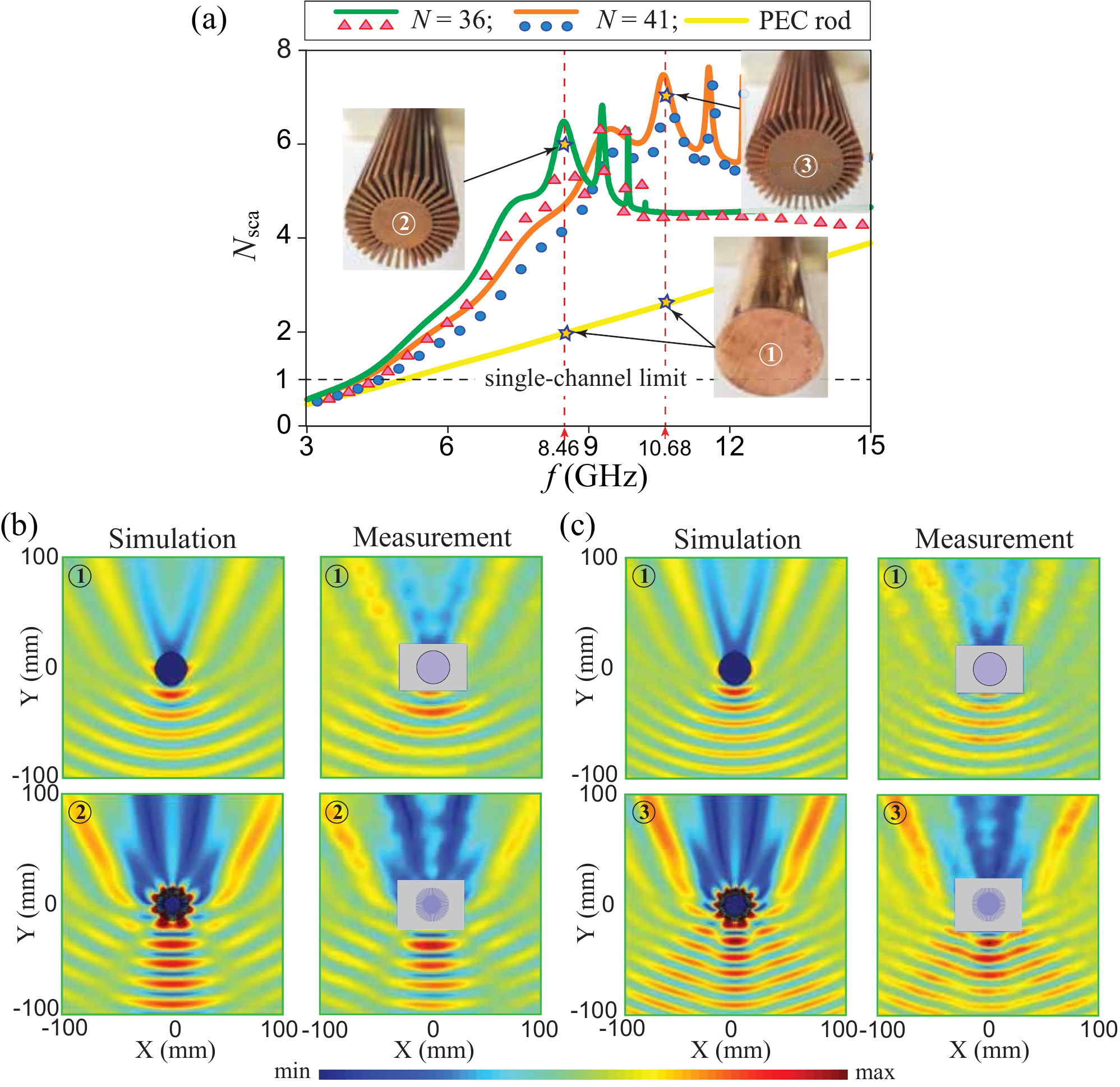}}
\caption{(a) Theoretical (lines) and simulated (markers) frequency dependence of scattering cross-sections of the TE$_z$ wave for a reference copper rod and two designed corrugated cylinders with $N=36$ and $N=41$, and patterns of the scattered electric near-field ($|E_x|$) simulated and measured at the frequencies of (b) 8.46 GHz and (c) 10.68 GHz. In the cases (b) and (c), the plane waves come from the bottom side and propagate along the $y$ axis.} \label{fig:fig6}
\end{figure*}

Figure \ref{fig:fig6}(a) shows theoretical and simulation results for the scattering cross-sections of the designed scatterers in the chosen frequency range. As is seen from this figure, there is some distinction between the theory and full-wave simulations. It is explained by coupling between azimuth spatial harmonics due to corrugations. This coupling is ignored in our theoretical model (see, Appendix \ref{sec:cylinder}) and increase in importance with increasing period of the corrugations \cite{tkachova_2019}. For the operating frequencies of designed superscatterers with 36 and 41 corrugations the ratio $p/\lambda$ is relatively large and approximately equal to 0.07 and 0.08, respectively. Despite this fact, for these frequencies our simple and fast theoretical approach still gives fairly accurate results.

To observe the superscattering from a corrugated cylinder the experimental sample was placed into an anechoic chamber. The sample is fixed vertically on the distance of $1.5$~m from a rectangular broadband horn antenna (HD-10180DRHA10SZJ). The antenna is connected to the first port of the R\&S{\textregistered}ZVA50 Vector Network Analyzer (VNA). The antenna generates a linearly polarized quasi-plane-wave whose polarization is orthogonal to the symmetry axis (the $z$ axis) of the sample. 

An electrically small dipole probe is positioned at the half height of the vertically standing scatterer and oriented along the polarization direction of the incident wave. The probe detects the dominant component ($E_x$) of the scattered electric near-field and is connected to the second port of the VNA. The LINBOU near-field imaging system is used for the near-field mapping. The scan area around the sample has the dimensions $200 \times 200$ mm in the $x$-$y$ plane. In the measurements, the probe automatically moves in the scan area with a 2 mm step along two orthogonal directions. At each probe position both amplitude and phase of the scattered electric field are measured. The exception is the area near the sample with dimensions $60 \times 44$ mm. This area is experimentally inaccessible because of the technical limitation on the distance between moving probe and vertically standing sample.

Figures \ref{fig:fig6}(b) and \ref{fig:fig6}(c) show the simulated and measured scattered electric near-field patterns ($|E_x|$) for a reference copper rod of the diameter of 3 cm (upper row) and two samples of the all-metal superscatterer (bottom row) at the operating frequencies. It can be seen that measurements are in reasonable agreement with simulations and indeed demonstrates the enhanced scattering of the TE$_z$ wave from the corrugated cylinder. These measurements present the first experimental evidence of superscattering for all-metal cylindrical objects. It should be noted that the designed scatterers feature relatively large transverse cross-sections. Therefore, they are characterized by the lower scattering efficiency than smaller scatterers [see, for instance, Fig.~\ref{fig:fig3}(a)]. The size of the superscatterer can be reduced by filling corrugations with dielectric. In practice, this, however, adds a complexity to the design and therefore calls for advanced fabrication techniques.

\section{Conclusions}
\label{sec:concl}

It is common knowledge that surface impedance offers a means for manipulating wave scattering from subwavelength objects. In this regard, the tensor impedance surfaces suggest the broadest potentials for application, but are now rare in practice because of the lack of feasible design solutions, especially in nonplanar geometries. It has been shown that metallic rod with helical periodic corrugations presents a concrete example of a cylindrical scatterer with the tensor impedance surface. Explicit expressions have been derived for amplitudes of an arbitrary-polarized plane waves scattered from such a cylinder. Particular attention has been given to the surface impedance, which enhances wave scattering from the cylinder, but induces no polarization conversion of incident wave. For the TE$_z$- and TM$_z$-polarized incident waves the required surface impedance of a subwavelength cylinder can be provided by longitudinal and transverse dielectric-filled corrugations, respectively. 

It has been demonstrated that such corrugations, when properly designed, initiate superscattering of waves, i.e. ensure resonance overlapping for two or more scattering modes (channels). This theoretical finding has been validated against full-wave simulations. The superscattering from a subwavelength corrugated cylinder takes place at multiple frequencies, which can be broadly tuned or even increased in number by adjusting the corrugation parameters. This phenomenon has been found to be robust to material losses and tolerances of corrugation fabrication, provided that the permittivity of corrugation filling is not too high. It has been shown that in this case enhanced scattering of the TE$_z$ waves from a cylinder features a broad frequency bandwidth in excess of 100\% with respect to the operating frequency. 

The broadest bandwidth of enhanced scattering can be achieved for hollow corrugations. In this case, the subwavelength corrugated cylinder behaves as a broadband all-metal superscatterer of the TE$_z$ waves. Two experimental samples of such superscatterers of the same radius have been designed and fabricated to operate in distinct frequencies. In the microwave band, the superscattering from these samples has been demonstrated by the electric near-field measurement, which presents the first experimental observation of this phenomenon for all-metal objects. 

It has been shown the the superscattering of the TM$_z$ waves from corrugated cylinder looks somewhat similar to that for the TE$_z$ waves. But, in addition to this effect, the corrugations are found to be responsible for cloaking of the cylinder under the incidence of the TM$_z$ waves. An efficient cloaking of subwavelength cylinder has been demonstrated as a result of simultaneous suppression of two lowest-order scattering modes. Besides, corrugated cylinder can be designed to exhibit cloaking at multiple frequencies. A unique feature of both multifrequency superscattering and multifrequency cloaking makes this scatterer particularly attractive for use in sensor and antenna applications. 

\section{Acknowledgement}
\label{sec:ack}

The authors are grateful for a support from Jilin University,
China.

\appendix

\section{Subwavelength corrugated cylinder}
\label{sec:cylinder}

Consider subwavelength periodic corrugations, which are parallel to the vector $\bf{i}_\parallel$ and have the period $p$ and the width $w$ along the $y$ axis and the depth $d$ along the $x$ axis [Figs.~\ref{fig:fig1}(b) and \ref{fig:fig1}(c)]. The corrugations are made of perfect electric conductor (PEC) and filled completely with dielectric of permittivity $\varepsilon$ and permeability $\mu=1$. Under the subwavelength condition, such corrugations do not support waves with nonzero component $E_\parallel$ of electric field \cite{harvey_1960, davies_1962, katsenelenbaum_1966, clarricoats_1969, scharten_1981, mahmoud_1991, iatrou_1996, barroso_1998, Tretyakov_2003}. Thus on the top (aperture) of corrugations one has the boundary condition $E_\parallel=0$, which is identical to that provided by a PEC surface $x=R$. By contrast, inside corrugations there are modes with nonzero electric field component $E_y$. Among them, fundamental mode with uniform field distribution along the $y$ axis is dominant for $w < p \ll \lambda$ and has the following field structure: $\{{\bf E},{\bf H}\}=\{{\bf E}(x),{\bf H}(x)\}\exp(-i\omega t + i k_\parallel x_\parallel)$, where $\omega$ and $\lambda=2\pi c/\omega$ are the mode frequency and wavelength, $x_\parallel$ and $k_\parallel$ are the coordinate and wavenumber along $\bf{i}_\parallel$, respectively. For this mode $E_y$ and $H_\parallel$ are related to each other by the expression:
\begin{equation}
E_y=-\frac{ik}{k^2_\bot}\frac{dH_\parallel}{dx},
\label{eq:electric}
\end{equation}
where $k^2_\bot=\varepsilon k^2-k^2_\parallel\approx\varepsilon k^2$ and the condition $k^2_\parallel \ll \varepsilon k^2$ is assumed to be fulfilled.

First, we consider rectangular corrugations [Fig.~\ref{fig:fig1}(b)]. For the fundamental mode of such corrugations one has the following solution of the wave equation: 
\begin{equation}
H_\parallel=A \cos[k_\bot(x-R_d)],
\label{eq:magnetic}
\end{equation}
which satisfies the PEC boundary condition on the bottom and side surfaces of corrugations for any unknown constant $A$.

Using Eqs.~(\ref{eq:electric}) and (\ref{eq:magnetic}), one can readily obtain the surface impedance along the corrugation width for $x=R$
\begin{equation}
W = \frac{E_y}{H_\parallel} = \frac{ik}{k_\bot}\tan(k_\bot d).
\label{eq:imped}
\end{equation}

Outside the corrugation aperture the condition $E_y/H_\parallel = 0$ holds true on the PEC surface $x=R$. The averaged boundary condition for the entire surface $x=R$ can be obtained by combining this condition with Eq.~(\ref{eq:imped}) and $E_\parallel(R)=0$
\begin{equation}
E_y = Z_\bot H_\parallel,~~~
E_\parallel = -Z_\parallel H_y,
\label{eq:system}
\end{equation}
where $Z_\bot=\langle E_y/H_\parallel \rangle = w/pW$ and $Z_\parallel=\langle E_\parallel/H_y \rangle = 0$ are the principal components of the surface impedance tensor
\begin{equation}
\hat{\bf Z}=\left(
\begin{matrix}
Z_\bot & 0 \\
0 & Z_\parallel
\end{matrix}\right),
\label{eq:matrix_z1}
\end{equation}
and parentheses $\langle \ldots \rangle$ denote averaging over $x=R$. Note that the obtained result, if necessary, can be generalized to account for finite conductivity of metallic corrugations \cite{shcherbinin_2017, shcherbinin_2019_2, tkachova_2019}. For good conductors the problem of wave attenuation due ohmic losses is of no concern in the microwave band, but is gaining in importance for the sub-terahertz and terahertz frequencies \cite{shcherbinin_2017, shcherbinin_2019_2, tkachova_2019, shcherbinin_2017_2}.

We now assume that the plane $x=r=R$ coincides with the surface of a cylindrical scatterer and orthogonal unit vectors ${\bf i}_z$ and ${\bf i}_\varphi$ of the cylindrical coordinate system are rotated by the angle $\theta$ with respect to the pair ${\bf i}_\parallel$ and ${\bf i}_y$ [Fig.~\ref{fig:fig1}(a)]. In the coordinates $\{r,\varphi,z\}$, the surface impedance tensor (\ref{eq:matrix_z1}) is generally non-diagonal 
\begin{equation}
\hat{\bf Z}=\left(
\begin{matrix}
Z_{\varphi\varphi} & Z_{\varphi z} \\
Z_{z \varphi} & Z_{zz}
\end{matrix}\right),
\label{eq:matrix_z2}
\end{equation}
and the boundary conditions on the cylinder surface $r=R$ (metasurface) take the form of Eq.~(\ref{eq:impedance}), where $Z_{\varphi \varphi}=Z_\bot\cos^2\theta + Z_\parallel\sin^2\theta$, $Z_{zz}=Z_\parallel\cos^2\theta + Z_\bot\sin^2\theta$, $Z_{\varphi z}=Z_{z \varphi} = (Z_\parallel - Z_\bot)\sin\theta \cos\theta$. 

Tensor (\ref{eq:matrix_z2}) describes the averaged surface impedance for a metallic cylinder with helical subwavelength corrugations. It takes the well-known diagonal form in the extreme cases of $\theta=0^\circ$ and $\theta=90^\circ$, which correspond to the PEC cylinder with longitudinal \cite{mahmoud_1991, iatrou_1996, shcherbinin_2017, shcherbinin_2019_2} and transverse \cite{harvey_1960, katsenelenbaum_1966, mahmoud_1991} rectangular corrugations, respectively. 

In the derivation of Eq.~(\ref{eq:matrix_z2}) we proceed from two assumptions. The first one is that the condition $k^2_\parallel \ll \varepsilon k^2$ is true, where $k^2_\parallel=(k_z\cos\theta + nR^{-1}\sin\theta)^2$ \cite{shcherbinin_2019_2, shcherbinin_2019}. For normal wave incidence the axial wavenumber $k_z$ equals zero and this condition reduces to the limitation $|n|\sin\theta\ll2\pi\sqrt{\varepsilon}R/\lambda$ on the azimuth index $n$ of modes in Eq.~(\ref{eq:azimuthal}). Simple estimation shows that, even with $\theta=90^\circ$, $\varepsilon=1$ and $R/\lambda\approx 1$, the first assumption in use is good for low azimuth harmonics with $n=0,\pm 1,\pm 2$.

The second widely-accepted assumption implies that the cylinder incorporates corrugations of rectangular shape \cite{katsenelenbaum_1966, iatrou_1996, dragone_1977}. This simplification takes no account of the cylinder curvature and is good for sufficiently large values of $k_\bot R_d$ (e.g., for large $\varepsilon$). One can avoid it in two practical cases. The first one is a metallic cylinder with longitudinal $(\theta=0^\circ)$ wedge-shaped corrugations \cite{davies_1962, scharten_1981, mahmoud_1991, barroso_1998}. In this case [Fig.~\ref{fig:fig1}(c)], substituting 
\begin{equation}
H_\parallel = H_z=A_1\left[J_0(k_\bot x) - BN_0(k_\bot x)\right]
\end{equation}
into Eq.~(\ref{eq:imped}) gives:
\begin{equation}
Z_{\varphi\varphi} = -\frac{w}{p}\frac{ik}{k_\bot}\frac{J^\prime_0(k_\bot R)-BN^\prime_0(k_\bot R)}{J_0(k_\bot R)-BN_0(k_\bot R)},~~~Z_{zz}=0,
\end{equation}
where $B=J^\prime_0(k_\bot R_d)/N^\prime_0(k_\bot R_d)$, $R_d=R-d$. Such anisotropic surface impedance has effect on the TE$_z$ waves \cite{shcherbinin_2015}.

The second case is a cylinder with transverse $(\theta=90^\circ)$ ring-shaped corrugations \cite{Barlow_1954, piefke_1959, harvey_1960, clarricoats_1969, mahmoud_1991, barroso_1998}, which affect normally incident TM$_z$ waves \cite{shcherbinin_2015}. In this case, using the field expressions 
\begin{equation}
E_z = A_2\left[J_n(k_\bot r) - CN_n(k_\bot r)\right],~~H_\varphi = \frac{i\varepsilon k}{k^2_\bot}\frac{dE_z}{dr},
\end{equation}
one obtains
\begin{equation}
Z_{\varphi\varphi} = 0,~~~Z_{zz} = \frac{w}{p}\frac{ik_\bot}{\varepsilon k}\frac{J_n(k_\bot R)-CN_n(k_\bot R)}{J^\prime_n(k_\bot R)-CN^\prime_n(k_\bot R)},
\label{eq:imped_zz}
\end{equation}
where $C=J_n(k_\bot R_d)/N_n(k_\bot R_d)$. Noteworthy is that Eq.~(\ref{eq:imped_zz}) depends on the azimuth index $n$. A similar situation holds for a dielectric cylinder coated by transverse (circumferential) \cite{sipus_2018} or helical \cite{shcherbinin_2018, shcherbinin_2019} conducting strips, as well as for a dielectric-lined metallic cylinder and multilayered-dielectric cylindrical structures \cite{mahmoud_1991, shcherbinin_2017_3}.

\bigskip

\bibliography{cylinder}

\end{document}